
\input phyzzx
%
\ifx\epsfbox\UnDeFiNeD\message{(NO epsf.tex, FIGURES WILL BE IGNORED)}
\def\figin#1{\vskip2.7in}
\else\message{(FIGURES WILL BE INCLUDED)}\def\figin#1{#1}\fi
%
\def\unlock{\catcode`@=11} 
\def\lock{\catcode`@=12} 
\unlock
\paperfootline={\hss\iffrontpage\else\ifp@genum\tenrm
    -- \folio\ --\hss\fi\fi}
\def\titlestyle#1{\par\begingroup \titleparagraphs
     \iftwelv@\fourteenpoint\fourteenbf\else\twelvepoint\twelvebf\fi
   \noindent #1\par\endgroup }
\def\GENITEM#1;#2{\par \hangafter=0 \hangindent=#1
    \Textindent{#2}\ignorespaces}
\def\papersize{\hsize=35pc \vsize=52pc \hoffset=0.5pc \voffset=0.8pc
   \advance\hoffset by\HOFFSET \advance\voffset by\VOFFSET
   \pagebottomfiller=0pc
   \skip\footins=\bigskipamount \normalspace }
\papers  
\def\address#1{\par\kern 5pt \titlestyle{\twelvepoint\sl #1}}
\def\abstract{\par\dimen@=\prevdepth \hrule height\z@ \prevdepth=\dimen@
   \vskip\frontpageskip\centerline{\fourteencp Abstract}\vskip\headskip }
\newif\ifYITP \YITPtrue
\font\fourteenmib =cmmib10 scaled\magstep2    \skewchar\fourteenmib='177
\font\elevenmib   =cmmib10 scaled\magstephalf   \skewchar\elevenmib='177
\def\YITPmark{\hbox{\fourteenmib YITP\hskip0.2cm
        \elevenmib Uji\hskip0.15cm Research\hskip0.15cm Center\hfill}}
\def\titlepage{\FRONTPAGE\papers\ifPhysRev\PH@SR@V\fi
    \ifYITP\null\vskip-1.70cm\YITPmark\vskip0.6cm\fi 
   \ifp@bblock\p@bblock \else\hrule height\z@ \rel@x \fi }
\def\schapter#1{\par \penalty-300 \vskip\chapterskip
   \spacecheck\chapterminspace
   \chapterreset \titlestyle{\ifcn@@\S\ \chapterlabel.~\fi #1}
   \nobreak\vskip\headskip \penalty 30000
   {\pr@tect\wlog{\string\chapter\space \chapterlabel}} }

\def\ssection#1{\par \ifnum\lastpenalty=30000\else
   \penalty-200\vskip\sectionskip \spacecheck\sectionminspace\fi
   \gl@bal\advance\sectionnumber by 1
   {\pr@tect
   \xdef\sectionlabel{\ifcn@@ \chapterlabel.\fi
       \the\sectionstyle{\the\sectionnumber}}%
   \wlog{\string\section\space \sectionlabel}}%
   \noindent {\S \caps\thinspace\sectionlabel.~~#1}\par
   \nobreak\vskip\headskip \penalty 30000 }

%
\newtoks\pubnum
\Pubnum={YITP/U-\the\pubnum}
\pubnum={9?-??}

%

%

%
\def\globaleqnumbers{\relax\if\equanumber<0\else\global\equanumber=-1\fi}
%
\def\addeqno{\ifnum\equanumber<0 \global\advance\equanumber by -1
    \else \global\advance\equanumber by 1\fi}
\mathchardef\Lag="724C  

%
 \def\ee{\eqno\eq }
 
 
 
 \def\overbar#1{\vbox{\ialign{##\crcr
           \vrule depth 2mm
           \hrulefill\vrule depth 2mm
           \crcr\noalign{\kern-1pt\vskip0.125cm\nointerlineskip}
           $\hfil\displaystyle{#1}\hfil$\crcr}}}


%
%

%
\def\sqr#1#2{{\vcenter{\hrule height.#2pt
      \hbox{\vrule width.#2pt height#1pt \kern#1pt
          \vrule width.#2pt}
      \hrule height.#2pt}}}


\def\Buildrel#1\under#2{\mathrel{\mathop{#2}\limits_{#1}}}
\def\llongrarrow{\hbox to 40pt{\rightarrowfill}}

\def\journals#1&#2(#3){\unskip; \sl #1~\bf #2 \rm (19#3) }
%
 \def\nllap#1{\hbox to-0.35em{\hskip-\hangindent#1\hss}}


%
\def\rslash{\partial\kern-0.026em\raise0.17ex\llap{/}%
          \kern0.026em\relax}
\def\Dslash{D\kern-0.15em\raise0.17ex\llap{/}\kern0.15em\relax}
\mathchardef\bigtilde="0365

%
\def\deqalign#1{\null\,\vcenter{\openup1\jot \m@th
    \ialign{\strut\hfil$\displaystyle{##}$&$\displaystyle{##}$&$
	\displaystyle{{}##}$\hfil\crcr#1\crcr}}\,}
%
%
%
\newcount\eqabcno \eqabcno=97 
\newcount\sequanumber \sequanumber=0
\newif\iffirstseq \firstseqtrue
\newif\ifequationsabc \equationsabcfalse
\def\eqnameabc#1{\relax\pr@tect
    \iffirstseq\global\firstseqfalse%
        \ifnum\equanumber<0 \global\sequanumber=\number-\equanumber
           \xdef#1{{\rm(\number-\equanumber a)}}%
           \global\advance\equanumber by -1
        \else \global\advance\equanumber by 1
           \xdef#1{{\rm(\ifcn@@ \chapterlabel.\fi \number\equanumber a)}}
        \fi
    \else\global\advance\eqabcno by 1
        \ifnum\equanumber<0
           \def#1{{\rm(\number\sequanumber \char\number\eqabcno)}}%
        \else
           \xdef#1{{\rm(\ifcn@@ \chapterlabel.\fi%
                        \number\equanumber \char\number\eqabcno)}}
        \fi
    \fi}

\def\eqsname#1{\relax\pr@tect%
        \ifnum\equanumber<0
           \def#1{{\rm(\number\sequanumber)}}
        \else
           \xdef#1{{\rm(\ifcn@@ \chapterlabel.\fi \number\equanumber)}}
        \fi}


\let\Title=\title  
\lock
\catcode`\@=11
%
%

%
%
%
\font\twelvemib   =cmmib10 scaled\magstep1	    \skewchar\twelvemib='177
\font\tenmib      =cmmib10			    \skewchar\tenmib='177
%
\font\fourteenbsy  =cmbsy10 scaled\magstep2     \skewchar\fourteenbsy='60
\font\twelvebsy    =cmbsy10 scaled\magstep1	      \skewchar\twelvebsy='60
\font\elevenbsy    =cmbsy10 scaled\magstephalf    \skewchar\elevenbsy='60
\font\tenbsy       =cmbsy10			      \skewchar\tenbsy='60
%
\newfam\mibfam
\def\mib{\n@expand\f@m\mibfam}
\let\tmpfourteenf@nts=\fourteenf@nts
\def\fourteenf@nts{\tmpfourteenf@nts %
    \textfont\mibfam=\fourteenmib      \scriptfont\mibfam=\tenmib
    \scriptscriptfont\mibfam=\tenmib }
\let\tmptwelvef@nts=\twelvef@nts
\def\twelvef@nts{\tmptwelvef@nts %
    \textfont\mibfam=\twelvemib	    \scriptfont\mibfam=\tenmib
    \scriptscriptfont\mibfam=\tenmib }
\let\tmptenf@nts=\tenf@nts
\def\tenf@nts{\tmptenf@nts %
    \textfont\mibfam=\tenmib   \scriptfont\mibfam=\tenmib
    \scriptscriptfont\mibfam=\tenmib }
%
\begingroup
\newcount\mibchar
\mibchar=\mibfam \multiply\mibchar by 256
\advance\mibchar by 11
\global\mathchardef\bfalpha  =\mibchar \advance\mibchar by 1
\global\mathchardef\bfbeta   =\mibchar \advance\mibchar by 1
\global\mathchardef\bfgamma  =\mibchar \advance\mibchar by 1
\global\mathchardef\bfdelta  =\mibchar \advance\mibchar by 1
\global\mathchardef\bfepsilon=\mibchar \advance\mibchar by 1
\global\mathchardef\bfzeta   =\mibchar \advance\mibchar by 1
\global\mathchardef\bfeta    =\mibchar \advance\mibchar by 1
\global\mathchardef\bftheta  =\mibchar \advance\mibchar by 1
\global\mathchardef\bfiota   =\mibchar \advance\mibchar by 1
\global\mathchardef\bfkappa  =\mibchar \advance\mibchar by 1
\global\mathchardef\bflambda =\mibchar \advance\mibchar by 1
\global\mathchardef\bfmu     =\mibchar \advance\mibchar by 1
\global\mathchardef\bfnu     =\mibchar \advance\mibchar by 1
\global\mathchardef\bfxi     =\mibchar \advance\mibchar by 1
\global\mathchardef\bfpi     =\mibchar \advance\mibchar by 1
\global\mathchardef\bfrho    =\mibchar \advance\mibchar by 1
\global\mathchardef\bfsigma  =\mibchar \advance\mibchar by 1
\global\mathchardef\bftau    =\mibchar \advance\mibchar by 1
\global\mathchardef\bfupsilon=\mibchar \advance\mibchar by 1
\global\mathchardef\bfphi    =\mibchar \advance\mibchar by 1
\global\mathchardef\bfchi    =\mibchar \advance\mibchar by 1
\global\mathchardef\bfpsi    =\mibchar \advance\mibchar by 1
\global\mathchardef\bfomega  =\mibchar\endgroup
%
%
\def\fourteenpoint{\fourteenf@nts \samef@nt \b@gheight=14pt \setstr@t }
\def\twelvepoint{\twelvef@nts \samef@nt \b@gheight=12pt \setstr@t }
\def\tenpoint{\tenf@nts \samef@nt \b@gheight=10pt \setstr@t }
\def\Tenpoint{\tenpoint\twelv@false\spaces@t}
\def\Twelvepoint{\twelvepoint\twelv@true\spaces@t}
\Twelvepoint  


\font\seventeentt=cmtt10 scaled\magstep3

\font\seventeenmib=cmmib10 scaled\magstep3  \skewchar\seventeenmib='177
\font\seventeenbsy=cmbsy10 scaled\magstep3   \skewchar\seventeenbsy='60
\font\seventeencp =cmcsc10 scaled\magstep3
\def\seventeenf@nts{\relax
    \textfont0=\seventeenrm         \scriptfont0=\twelverm
    \scriptscriptfont0=\ninerm
    \textfont1=\seventeeni          \scriptfont1=\twelvei
    \scriptscriptfont1=\ninei
    \textfont2=\seventeensy         \scriptfont2=\twelvesy
    \scriptscriptfont2=\ninesy
    \textfont3=\seventeenex     \scriptfont3=\seventeenex
    \scriptscriptfont3=\seventeenex
    \textfont\itfam=\seventeenit
    \textfont\slfam=\seventeensl \scriptfont\slfam=\twelvesl
    \textfont\bffam=\seventeenbf
    \scriptfont\bffam=\twelvebf  \scriptscriptfont\bffam=\ninebf
    \textfont\ttfam=\seventeentt
    \textfont\cpfam=\seventeencp
    \textfont\mibfam=\seventeenmib      \scriptfont\mibfam=\twelvemib
    \scriptscriptfont\mibfam=\tenmib }
\def\seventeenpoint{\seventeenf@nts \samef@nt \b@gheight=17pt \setstr@t }

\catcode`\@=12

\Pubnum={\the\pubnum}
\pubnum={UT-636\cr TIT/HEP-217\cr YITP/U-93-07\cr hep-th/9303123}
\date={March 1993}
\titlepage
\vskip13mm
\Title{\seventeenpoint\bf Ultraviolet Stable Fixed Point and Scaling
Relations in 2+$\bfepsilon$ Dimensional Quantum Gravity}
\vskip13mm
\author{
Hikaru Kawai \footnote*{{\twelvesl Department of Physics, University of
Tokyo, Hongo, Tokyo 113, Japan}
\quad\quad\quad E-mail address : TKYVAX\$hepnet::KAWAI},
Yoshihisa Kitazawa \footnote\dag{{\twelvesl Department of Physics, Tokyo
Institute of Technology, Oh-okayama, Meguro-ku, Tokyo 152, Japan }
\hfil\break E-mail address : TITVS0::KITAZAWA.decnet},
and Masao Ninomiya \footnote\#{{\twelvesl
Uji Research Center, Yukawa Institute for Theoretical Physics, Kyoto
University, Uji 611, Japan }
\hfil\break E-mail address : NINOMIYA@JPNYITP.bitnet}
}
\def\abstract#1{\vfill\vskip\frontpageskip\centerline%
               {\fourteenbf Abstract}\vskip\headskip#1\endpage}
\abstract{
We formulate a renormalizable quantum gravity in $2+\epsilon$ dimensions
by generalizing the nonlinear sigma model approach to string theory.
We find that the theory possesses the ultraviolet stable fixed point
if the central charge of the matter sector is in the range
$0~<~c~<~25$.
This may imply the existence of consistent quantum gravity
theory in 3 and 4 dimensions.
We compute the scaling dimensions of the relevant operators in the
theory at the ultraviolet fixed point.
We obtain a scaling relation between the cosmological constant and the
gravitational constant, which is crucial for searching for the continuum
limit in the constructive approach to quantum gravity.
}
\def\int{\intop\nolimits}
\overfullrule=0pt
\chapter{Introduction}

Recent development toward constructing a consistent theory of quantum
gravity has been pursued in two directions.
One is the string theory which has achieved a considerable success.
It turned out, however, that a nonperturbative treatment is needed for
further development.
Another approach is to describe the quantum gravity as a local field
theory.
It has been shown that the dynamical triangulation approach or matrix
model describes a consistent quantum gravity in two dimensions [1-3].
It was demonstrated that there exists a continuum theory of
quantum gravity in two dimensions which is renormalizable and
asymptotically free [4-7] and the exact scaling exponent can be
obtained [8-10].
The method of dynamical triangulation has been extended to 3 and 4
dimensional quantum gravity [11-13].

It is the purpose of the present article to study the continuum theory of
quantum gravity in $2+\epsilon$ dimensions.
In our previous work[14], we have formulated a quantum theory of Einstein
gravity in $2+\epsilon$ dimensions in such a way that the conformal
mode is explicitly separated.
The dynamics of the conformal mode is understood in terms of the
oversubtraction due to the one loop counter term.
The theory possesses the ultraviolet fixed point as long as the
central charge satisfies $0~<~c~<~25$, which separates weak and strong coupling
phases.
The exact scaling exponents of the two dimensional quantum gravity are
reproduced in the strong coupling regime when we take
$\epsilon \rightarrow 0$ limit.
In this paper we shall study the theory around the ultraviolet fixed
point as well as the weak coupling regime by making use of the renormalization
group.

In order to pursue further study of the dynamical triangulation,
we may need to know phase structures and the way how to
take the continuum limit.
We believe that these questions can be answered at least qualitatively
in the continuum theory by using the renormalization group.
In the present article we shall elucidate the properties of the phases and
the scaling relations among various operators.

This paper is organized as follows.
In section 2, quantum gravity in $2+\epsilon$ dimensions is formulated
in such a way that the conformal mode is explicitly separated.
A summary of our previous work is given.
In section 3, to study short distance properties of space time we
consider the most general renormalizable action and compute the one-
loop counter term.
In section 4, the renormalization group equations are derived and the
fixed point is found to be ultraviolet stable.
In section 5, the renormalization of the gravitationally dressed
operators is studied and their anomalous dimensions are computed
by utilizing the conformal invariance arguments in the nonlinear
sigma model approach to string theory.
It is pointed out that at the fixed point the target space is
de Sitter space.
We derive a scaling relation between the gravitational constant and
the cosmological constant. In section 6 we study two
dimensional gravity and string theory within our framework.
Section 7 is devoted to the conclusions and discussion.
There we discuss symmetries and renormalizability of the most general
renormalizable action.
\chapter{Formulation of $2+\varepsilon$ dimensional quantum
gravity and one loop renormalization}

In this section we summarize our previous work [14] for completeness.

\section{"Conformal gauge" formulation
in $2+\epsilon$ dimensions}

We adopt the following parametrization of $g_{\mu\nu}$
which separates the conformal mode from $h_{\mu\nu}$ :
$$ \eqalign{
 g_{\mu\nu}
       & = {\hat g}_{\mu\rho} {(e^h)}^\rho\,_\nu ~ e^{-\phi}\quad , \cr
       & = {\tilde g}_{\mu\nu} e^{-\phi}{~}_{,} \cr
}\ee $$
where ${\hat g}_{\mu\nu}$ is a background metric and $h^\mu_{~\nu}$
field is
taken to be traceless ${h^\mu}_\mu =0$, while $h_{\mu\nu}$ is
symmetric in $\mu$ and $\nu$.
Here tensor indices are raised and lowered by the background metric
${\hat g}_{\mu\nu}$.\footnote{*)}{In this article we use the notation of 't
Hooft and Veltman [15].}

The Einstein action in this parametrization is given by
$$
 {{\mu^\epsilon} \over G}\int d^D x \sqrt{g} R   =
   {{\mu^\epsilon} \over G} \int d^D x \sqrt{\hat g} ~
   e^{-{\epsilon\over2}\phi}
    \left\{ {\tilde R} - {1 \over 4} \epsilon
    (D-1) {\tilde g}^{\mu\nu} \partial_\mu \phi
   \partial_\nu \phi  \right\} ~.
	 \ee
$$
We change the integration variables from $g_{\mu\nu}$ to $h_{\mu\nu}$
and $\phi$.
The Jacobian for the change of variables is trivial with the
dimensional regularization.

We adopt the background field method to compute the quantum corrections
with the following gauge :
$$
{{\mu^\epsilon} \over G}\int d^D x {\sqrt{\hat g}} ~ {1\over2}
  \left({h^\nu}_{\mu,\nu} + {\epsilon\over2}\partial_\mu \phi\right)
  \left({h^{\rho \mu}}_{,\rho} + {\epsilon\over2}\partial^\mu
                                                      \phi\right) ~.
  \ee
$$
Expanding the action $(2.2)$ in terms of ${h^\mu}_\nu$ and $\phi$ and
adding $(2.3)$, we obtain the following quadratic action:
$$
\eqalign{
 {{\mu^\epsilon} \over G}\int d^D x {\sqrt{\hat g}} ~
	&  \biggl\{
	  {1\over 4}{h^\rho}_{\mu , \nu} {{h^\mu}_{\rho ,}}^\nu
   +{1\over 2}{\hat R}^\sigma\,_{\mu\nu\rho} {h^\rho}_\sigma
		h^{\mu\nu} \cr
	& -{\epsilon\over 8}D {\hat g}^{\mu\nu}
		\partial_\mu \phi \partial_\nu \phi
	+ {\epsilon\over 2}\phi {h^\mu}_\nu {\hat R}^\nu\,_\mu
	+ {{\epsilon^2}\over 8} \phi^2 {\hat R}
					\biggr\} ~. \cr
}\ee
$$
The interaction vertices are
$$
\eqalign{
{{\mu^\epsilon} \over G}\int d^D x \sqrt{{\hat g}} ~ & \biggl\{
{1\over 8} \epsilon^2
(D-1) \phi {\hat g}^{\mu\nu} \partial_\mu \phi \partial_\nu
\phi - {1\over 32}\epsilon^3 (D-1) \phi^2 {\hat g}^{\mu\nu}
\partial_\mu \phi \partial_\nu \phi \cr
& + {1\over 4} \epsilon (D-1) h^{\mu\nu} \partial_\mu \phi
\partial_\nu \phi - {1\over 8} \epsilon (D-1) {h^\mu}_\rho
h^{\rho\nu} \partial_\mu \phi \partial_\nu \phi \cr
& + \cdots  \biggr\} ~. \cr
}\ee $$
The ghost action is
$$
{{\mu^\epsilon}\over G} \int d^D x \sqrt{\hat g}
\left\{ {\bar\eta}^\mu{{{\eta_\mu},}^\nu}_\nu
                               - {\bar\eta}^\mu {\hat R}
^\nu_\mu \eta_\nu - {\epsilon \over 2} (\partial^\nu \phi )
{{\bar \eta}^\mu ,}_ \mu \eta_\nu + \dots \right\} \quad ~. \ee
$$
In the kinetic term of ${h^\mu}_\nu$ in (2.4), we expand the
background metric around the flat metric
$$
 {\hat g}_{\mu\nu} = \delta_{\mu\nu} + {\hat h}_{\mu\nu} \quad .
\ee
$$
The propagators are given by
$$
 \eqalign{ <h_{\mu\nu}(P) h_{\rho\sigma} (-P)> & =
{G\over P^2}
(\delta_{\mu\rho} \delta_{\nu\sigma} + \delta_{\mu\sigma}
  \delta_{\nu\rho}-{2\over D} \delta_{\mu\nu}
   \delta_{\rho\sigma}) \cr
<\phi (P) \phi(-P)> & = - {G\over P^2}
{4 \over {\epsilon D}} ~, \cr
<\eta^\mu (P){\bar \eta}_\nu(-P)> & = {G\over P^2}{{\delta^\mu}_\nu} .
                    \cr } \ee
$$

The propagator of the $\phi$ field has a $1\over\epsilon$
singularity which may be called the kinematical pole in the sense that
it does not arise from the high momentum part of the loop integral.

\section{One-loop renormalization and dynamics of
conformal mode}

The one-loop divergences of the theory including matter fields with
the central charge c come from the two types of graphs :
the tadpole graphs of ${h^\mu}_\nu$ and $\eta_\mu$ fields and the
well-known conformal anomaly graphs due to $\phi$,
${h^\mu}_\nu$, $\eta_\mu$ and the matter fields.
The total one-loop divergence is given by
$$
{{25-c}\over{24\pi}} {1\over \epsilon} \mu^\epsilon \int {d^D} x
\sqrt{\hat g}{\hat R}\quad .
\ee
$$
The theory is renormalizable with the one-loop bare action
$$
\mu^\epsilon \left({1\over G}-{{25-c}\over{24\pi}}
{1\over \epsilon}\right)   \int {d^D} x \sqrt{g} R
= {1\over G_0}\int {d^D} x \sqrt{g} R  \ee
$$
with the bare coupling
$$
{1\over G_0} = \mu^\epsilon \left({1\over G}-{{25-c}\over{24\pi}}
{1\over \epsilon}\right) . \ee
$$
The $\beta$~function is given by
$$
 \beta (G) = \epsilon G - {{25-c}\over{24\pi}} G^2 ~. \ee
$$
The ultraviolet fixed point $G^\ast$ determined by the condition
$\beta(G^\ast) =0$
is found to be
$$
G^\ast = {{24\pi} \over {25-c}} \epsilon \quad , \ee
$$
which exists as long as $c<25$.

We now consider the following one-loop bare action $(2.10)$ expressed in terms
of the parametrization $(2.1)$:
$$
\mu^\epsilon \int {d^D} x \sqrt{\hat g}
\left({1\over G}-{{25-c}\over{24\pi}}{1\over \epsilon}\right)
\left({\hat R} + {1 \over 4} {{h^\rho}_{\mu,\nu}{h^\mu}_\rho,}^\nu
+ {1 \over 2} {{\hat R}^\sigma}_{\mu\nu\rho}
{h^\rho}_\sigma h^{\mu\nu} + \dots \right)
$$
$$
- \mu^\epsilon \int {d^D} x \sqrt{\hat g}
\left({\epsilon\over {4G}} - {{25-c} \over {96\pi}}\right)
\left({\hat g}^{\mu\nu} \partial_\mu \phi \partial_\nu \phi +
2 \phi {\hat R} + \dots \right)~.  \ee
$$
The counter term of the $h_{\mu\nu}$ field is appropriate to make
the theory finite.
On the other hand, for the conformal mode $\phi$ there is
no divergence at the one-loop level.
In fact at the one-loop level the quantum correction to the kinetic
term of the conformal mode is of $O(\epsilon)$, since there is an
$\epsilon$ factor for each $\phi$ field.
Therefore the subtraction for $\phi$ in
$(2.14)$,
$$
{25-c \over 96\pi}\mu^\epsilon \int d^D x \sqrt{\hat g}
{\hat g}^{\mu\nu} \partial_\mu \phi \partial_\nu \phi \quad , \ee
$$
is an oversubtraction (it is of order $\epsilon^0$).
Thus $h_{\mu\nu}$ and $\phi$ are renormalized in a very
different manner.

We shall show in the following subsection that the use of the bare
coupling $G_0$ takes care of the oversubtraction problem in the
operator renormalization.
In particular, it allows us to consider the
$\epsilon \rightarrow 0$ limit, that is, the two-dimensional gravity.

Therefore the effective action for the conformal mode is given by
$$
- {\epsilon\over {4{G_0}}} \int d^D x \sqrt{\hat g} ( {\hat g}^{\mu\nu}
\partial_\mu \phi \partial_\nu \phi + 2 \phi {\hat R} )
+\dots \ee
$$
Note that when $G \gg \epsilon$ it is the Liouville action.

\section{Operator renormalization}

We consider the gravitational dressing of the operator
$$
\int d^D x {\sqrt{g}}^{1- {\Delta_0}} \Phi_{2{\Delta_0}} \quad , \ee
$$
where $\Phi_{2{\Delta_0}}$ is the spinless field with the scaling
dimension $2{\Delta_0}$.
In this subsection as well as the following one, we restrict our
consideration to the  strong coupling regime $G \gg \epsilon$
in which the bare coupling is given by
$$
{1\over G_0} = - {{25 -c} \over {24 \pi}}
{{\mu^\epsilon} \over \epsilon} \quad . \ee
$$
The perturbative calculation has been performed up to two-loop level
and the renormalization of the operator is given by
$$
Z_{\Delta_0} {\sqrt{g}}^{1-{\Delta_0}}
= Z_{\Delta_0} {\sqrt{\hat g}}^{1-{\Delta_0}}
  e^{{-{D\over 2}}(1-{\Delta_0}) \phi} \ee
$$
with
$$
\eqalign{
 Z_{\Delta_0} &= 1-\left({{G_0 \mu^\epsilon}\over{2\pi \epsilon}}
                                        (1 - {\Delta_0})\right)
     \left({1\over \epsilon} + \hbox{const.}\right)
      +{1\over 2}\,{1\over \epsilon}
        {\left({{G_0 \mu^\epsilon}\over{2\pi \epsilon}}
           (1- {\Delta_0})\right)}^2 \cr
   &  +{1\over 2}
     {\left({{G_0 \mu^\epsilon}\over{2\pi \epsilon}}
                 (1 - {\Delta_0}) \right)}^2
        {\left({1\over \epsilon} + \hbox{const.}\right)}^2
\quad . \cr
}\ee $$
The anomalous dimension follows is then given by
$$
\eqalign{
 \gamma_{{}_{\Delta_0}} & = \mu {\partial \over {\partial \mu}}
                     \log \, Z_{\Delta_0}   \cr
                  &  = -{{G_0 \mu^\epsilon}\over{2\pi \epsilon}}
                      (1 - {\Delta_0})
                   + {\left({{G_0 \mu^\epsilon}\over{2\pi \epsilon}}
                      (1 - {\Delta_0}) \right)}^2
                       + O ({G^3}_0 ) \quad . \cr
} \ee $$
Inserting the expression (2.18) for $G_0$ in the strong coupling
regime, we obtain
$$
 \gamma_{{}_{\Delta_0}} = {{4 (1-{\Delta_0)}}\over {Q^2}}
+ {\left({4 (1-{\Delta_0)}\over {Q^2}} \right)}^2 +
O \left({({4\over {Q^2}})}^3 \right)
\ee
$$
where
$$
Q = \sqrt{{25-c}\over3} \quad .
$$

We compare the ratio of the scaling exponents defined by
$$
{\beta\over \alpha} = {2(1-\Delta_0)+\gamma_{{}_{\Delta_0}}\over
{2+\gamma_{{}_{\Delta_0 = 0}}}}  \ee
$$
with that of the exact solutions of the two-dimensional gravity [8-10]
$$
\eqalign{\alpha & = -{Q\over 2} \left\{
{1-\sqrt{1-{8\over {Q^2}}} } \right\} \cr
\beta & = -{Q\over 2} \left\{
{1-\sqrt{1- {{8(1-\Delta_0)}\over Q^2}}
} \right\} ~.  \cr} \ee
$$
Apparently these two quantities agree up to the order of
${({4\over {Q^2}})}^2$ .

\section{Exact solution in the two-dimensional limit}

In the calculation of the operator renormalization (2.19) in the
strong coupling regime the loop graphs containing graviton
$h_{\mu\nu}$ are finite after cancellation of divergent parts.
Thus the $h_{\mu\nu}$ field does not play any role and we may drop
$h_{\mu\nu}$ in the one-loop bare action (2.10):
$$
{{\mu^\epsilon} \over G_0} \int d^D x \sqrt{g} R
\cong {{\mu^\epsilon} \over G_0} \int d^D x \sqrt{\hat g}
\left\{ {\hat R} e^{-{\epsilon\over 2}\phi}
-{\epsilon \over 4}(D-1) \sqrt{\hat g} ~
e^{-{\epsilon\over 2} \phi}
{\hat g}^{\mu\nu} \partial_\mu \phi \partial_\nu \phi \right\}~.
\ee
$$
In terms of a new variable $\psi$ introduced through
$$
e^{-{\epsilon \over 4}\phi} = 1 + {\epsilon \over 4} \psi \quad,
\ee
$$
the bare action becomes
$$
{{\mu^\epsilon} \over G_0} \int d^D x \sqrt{\hat g}
\left\{ {\hat R} {(1 + {\epsilon \over 4} \psi)}^2
- {\epsilon \over 8}D {\hat g}^{\mu\nu}
\partial_\mu \psi \partial_\nu \psi \right\} ~. \ee
$$

\noindent
The change of variables in (2.26) gives only the trivial Jacobian
with the dimensional regularization.

Inserting the expression of $G_0$ (2.18) into (2.27), we obtain the
following effective action for the $\psi$ field:
$$
 \eqalign{
 &  - {{25-c}\over{24\pi}}
{{\mu^\epsilon}\over \epsilon} \int d^D x\sqrt{\hat g} {\hat R}
{(1 + {\epsilon \over 4}\psi)}^2 \cr
& + {{25-c}\over{96\pi}} {D \over 2} \mu^\epsilon \int d^D x
\sqrt{\hat g} {\hat g}^{\mu\nu} \partial_\mu \psi
\partial_\nu \psi  ~ . \cr}
\ee
$$
This is a free theory with the propagator
$$
<\psi(P) \psi(-P)> = {{24\pi}\over{25-c}} {4\over D}
{1\over P^2} \quad .
\ee
$$

Therefore in performing the renormalization of the operator
$ \int d^D x {\sqrt{g}}^{1 - {\Delta_0}} \Phi_{2{\Delta_0}} $,
we may apply Wick's  contraction theorem.
To calculate expectation value
$$
<{\sqrt{g}}^{1-{\Delta_0}}> = \sqrt{\hat g}^{1-\Delta_0}
< \hbox{exp} ~ \left\{
{4\over \epsilon} (1-{\Delta_0}) \hbox{log}
(1 + {\epsilon \over 4}\psi) \right\} >  \ee
$$
we associate each contraction with a divergent factor of
$$
< \psi^2 > = - {24\over{25-c}} {1\over \epsilon} + \hbox{const} ~.
\ee
$$
The expectation value (2.30) can be evaluated exactly by using a
zero-dimensional model with the action
$$
{1\over 2} {{c-25}\over 24} \epsilon {\psi}^2  ~.
\ee$$

Thus (2.30) reduces to the ordinary integral in terms of a new
variable $\rho = {\epsilon\over 4 }\psi$:
$$
<{\sqrt{g}}^{1-\Delta_0}> = {\sqrt{g}}^{1-\Delta_0}
                            \hbox{const.} \int^\infty_{-\infty}d \rho ~
 \hbox{exp} \left\{ {4\over \epsilon} (1 - \Delta_0)
                   \hbox{log} (1 + \rho)
              -{1\over \epsilon} {{c-25}\over 3} \rho^2
            \right\} ~.  \ee
$$
By means of the saddle point method with the saddle point
$$
\rho_0 = {1\over 2}
\left\{-1 +  \sqrt{ 1- {8\over{Q^2}}(1 - \Delta_0)}\right\},
\ee
$$
the integration in (2.33) is evaluated to be
$$
 \hbox{exp} \lbrace
{4\over \epsilon} (1 - \Delta_0) \log (1 + \rho_0)
+ {Q^2 \over \epsilon}  {\rho_0}^2 \rbrace ~.
\ee
$$
The renormalization constant of the operator
$Z_{\Delta_0} {\sqrt{g}}^{1-\Delta_0}$ is given by
$$
Z_{\Delta_0} = \hbox{exp} \left\{
-{4\over \epsilon} (1 - \Delta_0) \log (1 + \rho_0)
- {Q^2 \over \epsilon} {\rho_0}^2 \right\} ~.
\ee
$$
The anomalous dimension is found to be
$$
 \gamma_{\Delta_0} = \mu {\partial\over {\partial\mu}}
                          {\rm log}Z_{\Delta_0}
                   = {\rho_0}^2 Q^2
                   = -2(1 - \Delta_0) - Q^2 \rho_0 ~.  \ee
$$
The ratio of the scaling exponents
${\beta \over \alpha} =
{{2(1-{\Delta_0}) + \gamma_{\Delta_0}} \over
                                     {2 + \gamma_{{\Delta_0} = 0}}}$
reproduces the exact solution of two-dimensional gravity given by
(2.24).

\chapter{One-loop divergence
                        in the most general renormalizable action}

In the preceding section the one-loop bare action (2.10) for the
Einstein action was derived.
As was stressed, $h_{\mu\nu}$ and conformal mode $\phi$
or $\psi$ defined\footnote{*)}{Note that hereafter we use a
definition different from that of section 2,
$e^{-{\epsilon \over 4} \phi}=1+{\epsilon \over 4}\psi$. } by
$e^{-{\epsilon\over 4} \phi}=1+{1\over 2} \sqrt{\epsilon\over 2} \psi$
are renormalized in a very different manner.
In particular the coefficient $-{\epsilon\over{2G_0}}$ of the kinetic term
(2.14) for the $\phi$ field formally vanishes around the ultraviolet
fixed point given by (2.13).
Thus we need to take into account the $O(\epsilon)$
contribution to the one-loop renormalization of the $\phi$ field.

In order to study the dynamics of the conformal mode around the
ultraviolet fixed point, we take a different approach from that of
section 2. If we retain the form
 of the Einstein action
and subtract the counter term, we would inevitably encounter the
oversubtraction problem of the $\phi$ field as was argued in
subsection 2.2.
For the purpose of avoiding the oversubtraction problem, we relax the
functional form of the $\phi$ field in the action and consider the
most general renormalizable action.

In two dimensions $\phi$ is dimensionless so that the condition of
renormalizability does not specify the form of the action.
In fact any function of $\phi$ is allowed.
The most general renormalizable action including the matter fields
$\varphi^i \, (i=1, \dots, c)$ with the central charge c takes the
form
$$
\int d^D x {\sqrt {\hat g}}
\left\{ {{\mu^\epsilon}\over G}{\tilde R} L(\psi) -
                                          {{\mu^\epsilon} \over {2G}}
      {\tilde g}^{\mu\nu} \partial_\mu \psi \partial_\nu \psi
    + {1\over 2} \sum^c_{i=1} {\tilde g}^{\mu\nu}
      \partial_\mu \varphi^i
        \partial_\nu \varphi^i e^{\Phi(\psi)} \right\}~~, \ee
$$
where $\tilde{R}$ is the scalar curvature constructed out of
${\tilde g}_{\mu\nu}= {\hat g}_{\mu\rho}{(e^h)}^\rho_\nu$ .
Here we impose the following conditions:
$$
L(0)=1 \quad ,\Phi(0)=0 \quad .\ee
$$
We have defined the $\psi$ field in such a way that the kinetic term
becomes canonical (see $(3.1)$).
We have assumed that the matter coupling term is invariant under
$\varphi^i \rightarrow \varphi^i + {\rm const}$ .
The functional $L(\psi)$ and $\Phi(\psi)$ are left arbitrary unless
some symmetry is imposed on $(3.1)$.
For further discussion on symmetries, see section 7.

We believe that the long distance property of space time is well described
by
the theory with the Einstein action.
This is also required from the general covariance.
Therefore the general covariance is imposed by
demanding that the most general renormalizable action $(3.1)$
coincides with the Einstein action in the infrared limit
$\mu \sim 0$ where the coupling $G$ becomes weak.
In terms of $\psi$ the Einstein action reads
$$
\eqalign{
\int d^D x {\sqrt {\hat g}}
\biggl\{ & {{\mu^\epsilon}\over G}{\tilde R}
{\left(1 + {1\over 2} {\sqrt {\epsilon\over 2}} \psi \right)}^2
- {{\mu^\epsilon}\over 2G} {\tilde g}^{\mu\nu} \partial_\mu \psi
                                              \partial_\nu\psi\cr
&+ {1\over 2}
       {\left(1 + {1\over 2} {\sqrt {\epsilon\over 2}} \psi \right)}^2
  {\tilde g}^{\mu\nu}
                \partial_\mu \varphi^i \partial_\nu \varphi^i \biggr\}.
}\ee $$
Therefore in the weak coupling limit $G \sim 0$, or $\mu \sim 0$,
$L(\psi)$ and $\Phi(\psi)$ must behave as
$$
\eqalignno{
& L(\psi) \sim 1 + {\sqrt{\epsilon \over 2}} \psi \cr
&\Phi(\psi) \sim {\sqrt{\epsilon \over 2}} \psi \quad . &\eq\cr
}$$

We are going to study the short distance property of space time in the
following method.
First we calculate the one-loop bare action for the most general
renormalizable action and then derive renormalization group equations
for G,$L(\psi)$ and $\Phi(\psi)$.
While G,$L(\psi)$ and $\Phi(\psi)$ evolve according to the $\beta$
functions by a change of the renormalization scale $\mu$, these
trajectories are specified by the requirement that the theory
coincides with the Einstein action in the infrared limit.
As will be shown in section 4 the $\beta$ functions possess an
ultraviolet stable fixed point and thus the short distance property of
the theory is governed by this fixed point.
The action at this point is determined from the renormalization group
equations.

In the rest of this section we compute the one-loop divergence of the
action $(3.1)$ by the background field method.
To this end we make a shift of variables:
$$
\eqalignno{
\psi & \rightarrow {\bar \psi} + \psi \cr
\varphi^i & \rightarrow {\bar \varphi}^i + \varphi^i &\cr
}$$
where on the right hand side ${\bar \psi}$ and ${\bar \varphi}^i $
are the background fields while $\psi$ and $\varphi^i$ denote the
quantum fields.
The gauge fixing term is chosen to be of a similar form to $(2.3)$:
$$
{{\mu^\epsilon} \over G} \int d^D x {\sqrt{\hat g}}
{1\over 2} L \left({\bar \psi} \right)
{\left\{ {h^\nu}_{\mu , \nu} -
\partial_\mu \left({{L \left( {\bar\psi} + \psi \right)} \over
{L(\bar\psi)}}\right)
\right\}}^2. \ee
$$
The ghost action is then given by
$$
{{\mu^\epsilon} \over G} \int d^D x {\sqrt{\hat g}}
L \left({\bar \psi} \right)
\biggl\{{\bar\eta}^\mu {{\eta_\mu,}^\nu}_\nu
                  - {{\hat R}^\nu}_{~\mu} {\bar\eta}^\mu \eta_\nu
                  - \bar\eta^\mu
                     \Bigl({1\over {L({\bar\psi})}}
                    \partial^\nu L \left( {\bar\psi} + \psi \right)
                    \eta_\nu \Bigr)_{,_\mu} + \dots \biggr\}.
\ee
$$

In order to evaluate one-loop divergences we expand $(3.1)$ together
with $(3.5)$ in terms of the $h_{\mu\nu}$ , $\psi$ and $\varphi^i$
fields with the help of the formula:
$$
\eqalign{{\tilde R} &
= {\hat R} - {h^\mu}_\nu {{\hat R}^\nu}_{~\mu} - {h^{\mu\nu},}_{\mu\nu}
  + {1\over 4} {h^\mu}_{\nu,\rho} {h^\nu}_\mu,^\rho \cr
& + {1\over 2} {{\hat R}^\sigma}_{\mu\nu\rho} {h^\rho}_\sigma
                                                         h^{\mu\nu}
  - {1\over 2} {h^\nu}_{\mu , \nu} {h^{\rho\mu}}_{,\rho}
  + {\left( {h^\mu}_\nu
                {{h^\nu}_{\rho ,}}^\rho \right),}_\mu
\cr
& + \dots \quad . \cr
}\ee$$
The tree action is
$$
\int d^D x {\sqrt{\hat g}}
\left\{\mu^\epsilon  {\hat R} {L\over G}
    - {\mu^\epsilon \over {2G}} \partial_\mu {\bar\psi}
                                             \partial^\mu {\bar\psi}
    + {1\over 2} e^\Phi \partial_\mu {\bar\varphi}^i
                        \partial^\mu {\bar\varphi}^i \right\} \ee
$$
where $L=L\left({\bar\psi}\right)$ and
$\Phi = \Phi\left({\bar\psi}\right)$ .
The kinetic terms are given by
$$
\int d^D x {\sqrt{\hat g}}
\biggl\{\mu^\epsilon {L\over {4G}} {h^\mu}_{\nu,\rho}
                                             {{h^\nu}_{\mu,}}^\rho
   -{\mu^\epsilon \over {2G}} \partial_\mu \psi \partial^\mu \psi
   +{1\over 2} e^\Phi \partial_\mu \varphi^i \partial^\mu \varphi^i
\biggr\}. \ee
$$
Other quadratic terms including the gauge fixing term $(3.5)$ are
$$
\eqalignno{
\int d^D x {\sqrt{\hat g}}&
\biggl\{
{\mu^\epsilon \over 2}{\hat R} {{L''} \over G} \psi^2
- \mu^\epsilon{{L'} \over G} {{\hat R}^\nu}_{~\mu} {h^\mu}_\nu \psi
+ \mu^\epsilon{L \over {2G}} {{\hat R}^\sigma}_{\mu\nu\rho}
{h^\rho}_\sigma  h^{\mu\nu}
+ \mu^\epsilon{L\over G} ({h^\mu}_\nu {{h^\nu}_\rho,}^\rho) ,_\mu \cr
&+ {1\over 4}e^\Phi \partial_\mu {\bar\varphi}^i
                                       \partial^\mu {\bar\varphi}^i
   \lbrack {(\Phi')}^2 + \Phi'' \rbrack \psi^2
+ \Phi'e^\Phi \partial_\mu {\bar\varphi}^i \partial^\mu \varphi^i \psi
\biggr\}. &\eq\cr
}$$

Before going into the evaluation of the divergence we make a redefinition
of the field variables
$$
\eqalignno{
L^{1\over 2} {h^\mu}_\nu & \rightarrow {h^\mu}_\nu \cr
e^{\Phi\over 2} \varphi^i& \rightarrow \varphi^i
&\eq\cr
}$$
in such a way that the kinetic terms become canonical.
Thus, by retaining only the terms relevant to the one-loop
divergences, eq. (3.9) reads
$$
\eqalign{
&\int d^D x {\sqrt{\hat g}}
\biggl\{ {\mu^\epsilon\over {4G}} {h^\mu}_{\nu,\rho}
                                                   {{h^\nu}_\mu,}^\rho
  - {\mu^\epsilon\over{2G}}\partial_\mu \psi \partial^\mu \psi
  + {1\over 2} \partial_\mu \varphi^i \partial^\mu \varphi^i \cr
&  +{\mu^\epsilon\over {16G}} {\left({{L'}\over L} \right)}^2
  \partial_\mu {\bar\psi} \partial^\mu {\bar\psi}
                                     {h^\rho}_\nu {h^\nu}_\rho
  + {1\over 8}{(\Phi')}^2 \partial_\mu {\bar\psi}
                                 \partial^\mu {\bar\psi}{(\varphi^i)}^2
  + \dots \biggr\} ~, \cr
}\ee$$
while other quadratic terms are
$$
\eqalign{
&\int d^D x {\sqrt{\hat g}}
\biggl\{ {\mu^\epsilon\over 2} {\hat R} {{L''}\over G} \psi^2
+ {\mu^\epsilon\over {2G}}{{\hat R}^\sigma}_{\mu\nu\rho}
                                   {h^\rho}_\sigma h^{\mu\nu}\cr
&+ {1\over 4} e^\Phi \partial_\mu {\bar\varphi}^i
                                     \partial^\mu {\bar\varphi}^i
   \lbrack {(\Phi')}^2 + \Phi'' \rbrack \psi^2
+  \Phi' e^{\Phi\over 2} \partial_\mu {\bar \varphi}^i
                                     \partial^\mu \varphi^i \psi
+ \dots \biggr\}. \cr
}\ee$$
In the ghost action (3.6) we also make a similar redefinition
$L \bar\eta_\mu \rightarrow \bar\eta_\mu$
to obtain
$$
{{\mu^\epsilon}\over G} \int d^D {\sqrt{\hat g}}
\biggl\{ {\bar\eta}^\mu {{\eta_\mu,}^\nu}_\nu
                  - {{\hat R}^\nu}_{~\mu} {\bar\eta}^\mu \eta_\nu
+\left( {{L'}\over L} \right) \partial_\nu {\bar\psi}
  \bigl({{\bar\eta}^\mu}_{~,\mu} \eta^\nu  \bigr)
 + \dots \biggr\}. \ee
$$

We now evaluate the one-loop divergences.
The kinetic terms in $(3.12)$ give divergences of the conformal
anomaly.
One-loop divergence due to a scalar field is
$$
- {1\over {24\pi \epsilon}} {\sqrt{\hat g}} {\hat R}.  \ee
$$
By counting the number of degrees of freedom for $h_{\mu\nu}$ and the
ghost as 2 and -~4 respectively, the total divergences coming from the
conformal
anomaly are given by
$$
{{1-c}\over {24\pi\epsilon}} {\sqrt{\hat g}} {\hat R}.  \ee
$$
The rest of the divergences in $(3.12)$ stems from the tadpole graphs
of the last two terms
$$
{\left( \Phi' \right)}^2
{\sqrt{\hat g}} {1\over {16G}} {\left({{L'}\over L}\right)}^2
\partial_\mu {\bar\psi} \partial^\mu {\bar\psi}
< {h^\rho}_\nu {h^\nu}_\rho >
= -{1\over {8\pi\epsilon}}{\left({{L'}\over L}\right)}^2
{\sqrt{\hat g}} \partial_\mu {\bar\psi} \partial^\mu{\bar\psi}~~, \ee
$$
and
$$
{\sqrt{\hat g}} {1\over 8} {\left( \Phi' \right)}^2
\partial_\mu {\bar\psi} \partial^\mu {\bar\psi}
< {\left(\varphi^i \right)}^2 >
= {-c \over {16\pi\epsilon}} {(\Phi')}^2 {\sqrt{\hat g}}
\partial_\mu {\bar\psi} \partial^\mu {\bar\psi}~~, \ee
$$
where the propagator in $(2.8)$ is used.
We next consider $(3.13)$.
The tadpole divergences arising from the 1st and 2nd terms are
$$
\eqalignno{
{1\over {4\pi\epsilon}} & {\sqrt{\hat g}} {\hat R} L'' ~~,  \cr
{1\over {2\pi\epsilon}} & {\sqrt{\hat g}} {\hat R} ~~, &\eq \cr}
$$
while the last two terms give altogether
$$
{G\over{8\pi\epsilon}}
\Phi'' e^\Phi {\sqrt{\hat g}} \partial_\mu {\bar\varphi}^i
                             \partial^\mu {\bar\varphi}^i . \ee
$$
In the ghost action $(3.14)$, the second term gives rise to the tadpole
divergence
$$
{1\over {2\pi\epsilon}}{\sqrt{\hat g}} {\hat R}  \ee
$$
and the rest of the terms give
$$
{1\over {8\pi\epsilon}}{\left({{L'}\over L} \right)}^2 {\sqrt{\hat g}}
   \partial_\mu {\bar \psi} \partial^\mu {\bar\psi}. \ee
$$
All other graphs constructed from $(3.12)$ , $(3.13)$ and $(3.14)$
are either finite or total derivatives.
Summing up the divergences $(3.16) - (3.22)$
we obtain the following one-loop counter term
$$
{\sqrt{\hat g}} \biggl\{
-{A\over \epsilon}{\hat R} - {1\over {4\pi\epsilon}}{\hat R} L''
+ {c\over {16\pi\epsilon}}{\left(\Phi'\right)}^2
  \partial_\mu {\bar\psi} \partial^\mu {\bar\psi}
- {G\over{8\pi\epsilon}} \Phi'' e^\Phi
\partial_\mu {\bar\varphi}^i \partial^\mu {\bar\varphi}^i \biggr\}~~,
\ee
$$
where
$$
A = {{25-c} \over {24\pi}} \quad .
$$

Note that $(3.17)$ and $(3.22)$ have cancelled each other.
Hence $h_{\mu\nu}$ and the ghost fields contribute to the well-known
 counter term $-{26 \over{24\pi\epsilon}}{\hat R}$ only.
\chapter{Renormalization Group and Fixed Points}

{}From the one-loop counter term in the previous
section, we can express the bare quantities in terms of the
renormalized quantities :
$$
{1 \over G_0} = {\mu^\epsilon}
{\left\{ {1 \over G} - {A \over \epsilon} -
{{L''(0)} \over {4 \pi \epsilon}} \right\}},
$$

$${1 \over G_0} \partial_\mu \psi_0 \partial^\mu \psi_0
= {\mu^\epsilon \over G} \partial_\mu \psi \partial^\mu \psi
{\left\{ 1 - {c \over {8 \pi \epsilon}} \left(\Phi'(\psi)\right)^2
 \right\}},
$$

$$
{1 \over G_0} L_0 (\psi_0) = \mu^\epsilon
    \left\{{1 \over G} L(\psi) -{{L''(\psi)} \over {4 \pi \epsilon}}
               - {A \over \epsilon} \right\} \quad ,
$$

$$
\varphi_0^i = \varphi^i \left\{ 1 - {G \over {8 \pi \epsilon}}
\Phi''(0) \right\},
$$

$$
\Phi_0 (\psi_0) = \Phi(\psi)-{G \over {4 \pi \epsilon}} \Phi''(\psi) +
{G \over { 4 \pi \epsilon}} \Phi''(0).
\ee
$$

Let us recall that the functions $L$ and $\Phi$ are defined such that
$L (0) = 1$ and $\Phi (0) = 0$ .
Therefore when we determine these relations, we have also required
that $L_0 (0) = 1$ and $\Phi_0 (0) = 0$.
{}From these relations, we find
$$
\psi_0 = \psi - G \int_0^\psi dy \left\{ {c \over {16 \pi \epsilon}}
{(\Phi ' (y) )^2} -
{A \over {2 \epsilon}} - {{L''(0)} \over {8 \pi\epsilon}} \right\},
\ee$$

$$
\eqalign{
L_0 (\psi_0) & = L(\psi) -{G \over {4\pi\epsilon}}
                 L''(\psi)+{{GA}\over \epsilon}L(\psi)+
                 {{GL''(0)} \over {4\pi\epsilon}} L(\psi)
                  - {{GA} \over \epsilon} ~~,\cr
             & = L(\psi_0) + L'(\psi_0) G
                 \int_0^{\psi_0} dy
                 \left\{ {c \over {16\pi\epsilon}}
                 {\bigg({\Phi'(y)}\bigg)}^2
                 - {A\over {2\epsilon}}
                 - {{L''(0)} \over {8\pi\epsilon}} \right\} \cr
             &~~~~~-{G \over {4 \pi \epsilon}}{L''(\psi_0)}
                 + {{GA} \over \epsilon}  L (\psi_0) +
                 {{GL''(0)} \over {4\pi\epsilon}} L(\psi_0)
                 - {{GA} \over \epsilon}~~,  \cr
 }$$

$$
\eqalign{
\Phi_0(\psi_0) & = \Phi(\psi) - {G \over {4\pi\epsilon}}
                   \Phi''(\psi)+{G \over{4\pi\epsilon}}\Phi''(0) ~~,\cr
               & = \Phi(\psi_0) + \Phi'(\psi_0) G \int_0^{\psi_0}
                   dy \left\{{c \over {16 \pi\epsilon}}
                   {\left({\Phi'} (y)\right)}^2
                   - {A \over {2\epsilon}}
                   - {{L''(0)} \over {8\pi\epsilon}} \right\} \cr
               &~~~~~ -{G \over {4\pi\epsilon}} \Phi''(\psi_0)
                   + {{G \over {4\pi\epsilon}} \Phi''(0)}. \cr
}\ee$$

The $\beta$~and $\gamma$~functions (functionals to be precise) follow
by demanding that the bare quantities are independent of $\mu$.
At the tree level we have
$$
\beta_G \equiv \mu{{dG}\over {d\mu}} = \epsilon G ,
$$
$$
\beta_L \equiv \mu {{dL} \over {d\mu}} = 0 ,
$$
$$
\beta_\Phi \equiv \mu {{d\Phi} \over {d\mu}} =0 ,
$$
$$
\gamma_\psi\equiv \mu{{d\psi} \over {d\mu}} = 0 ,
$$
$$
\gamma _{\varphi^i}\equiv \mu{{d\varphi^i}\over {d\mu}} = 0 .
\ee$$
By utilizing the tree level $\beta$~and $\gamma$~functions,
the one-loop level $\beta$~and $\gamma$~functions are found to be
$$
\beta_G = \epsilon G - \left(A + {{L''(0)} \over {4\pi}} \right) G^2 ,
$$
$$
\beta_L = -GL'(\psi) \int_0^\psi dy
         {\left\{ {c \over {16 \pi}} {\bigg({\Phi'(y)}\bigg)}^2
          - {A \over 2} - {{L''(0)}\over {8\pi}} \right\} }
$$
$$        + \left\{ A + {1\over {4\pi}} L''(\psi) - AL(\psi)
          - {L''(0) \over {4\pi}}L(\psi) \right\}   G,
$$
$$
\beta_\Phi = -G \Phi'(\psi) \int_0^\psi dy
      {\left\{ {c \over {16\pi}} {\left(\Phi'(y) \right)}^2
       - {A \over 2} - {{L''(0)} \over {8\pi}} \right\}}
$$
$$
      + {\left\{ {1\over {4 \pi}}\Phi''(\psi)  -
      {{\Phi''(0)} \over {4\pi}} \right\} } G \quad ,
$$
$$
\gamma_\psi = G\int_0^\psi dy
             \left\{ {c\over {16\pi}} {\bigg(\Phi'(y)\bigg)}^2
             -{A\over 2} - {{L''(0)} \over {8\pi}} \right\} \quad ,
$$
$$
\gamma_{\varphi^i} = {G\over {8\pi}} \Phi''(0) \varphi^i .
\ee$$

As we change the renormalization scale $\mu$, $G$, $L$ and $\Phi$
evolve
according to the $\beta$~functions.
The renormalization group trajectory is specified by the requirement
that the theory coincides with the Einstein action $(3.3)$
in the infrared
limit $\mu \rightarrow 0$ where the coupling constant G is weak.
The reason is that the full theory is given by the tree action when G
is weak and hence the general covariance of the theory can be ensured
by the general covariance of the tree action in the weak coupling
limit.
Therefore we choose the following initial conditions
$$
e^{\Phi(\psi)} {\biggl |}_{\mu\sim 0} = L(\psi) {\biggl |}_{\mu\sim 0}
=\, {\left(1 + {1 \over 2} {\sqrt{\epsilon\over 2}} \psi \right)}^2
\ee$$
and let the
functions evolve according to the $\beta$~functions as we increase
the renormalization scale $\mu$.
Since we have derived the renormalization group equations of the theory and the
way to specify its renormalization trajectory, we need to study the
property of this renormalization group and its ultraviolet fixed point
in particular.

It is easy to observe that the $\beta$~functions possess the following
fixed point:
$$
G^\ast = {{24\pi}\over {25-c}}\epsilon  \, ,
$$

$$
L^\ast (\psi) = 1 \, ,
$$

$$
\Phi^\ast (\psi) = \sqrt {{25-c}\over {3c}} \psi.
\ee$$
In order to study the physical property of this fixed point, we expand
the $\beta$~functions around the fixed point and obtain
$$
\beta_G = - \epsilon \delta G - {{{G^\ast}^2} \over {4 \pi}}
\delta L''(0),
$$

$$
\beta_L = \left\{ -A \delta L (\psi) +
{{\delta L''(\psi)} \over {4\pi}} - {{\delta L'' (0)} \over {4\pi}}
 \right\} G^\ast,
$$

$$
\beta_\Phi = \left\{ - {A \over 2} \delta \Phi (\psi) +
{{\delta \Phi'' (\psi)} \over {4\pi}}
- {{\delta \Phi''(0)} \over {4\pi}} \right\} G^\ast + G^\ast \Phi^\ast
{{\delta L''(0)} \over {8\pi}}~~,
\ee$$
where $\delta G$, $\delta L$ and $\delta \Phi$ are the deviations from
the fixed point.
In order to find the eigenvalues and eigenfunctions of these
linearized $\beta$~functions we need to specify the physically
acceptable function space.
We may require that the functions $L(\psi)$ and $\Phi (\psi)$ do not
grow exponentially at infinity $(\vert \psi \vert \rightarrow \infty)$
{}.
Furthermore  $\delta L(0) = \delta \Phi (0) = 0$.
Then the eigenfunctions of the linearized $\beta$~functions and
their eigenvalues are
$$\delta L = \sin \Lambda \psi, $$
$$\delta \Phi = \sin \Lambda ' \psi, $$
$$\beta_G = - \epsilon \delta G, $$
$$\beta_L = -(A + {\Lambda^2 \over {4\pi}}) G^\ast \delta L,$$
$$\beta_\Phi =
           - \left( {A \over 2} + {{\Lambda'}^2 \over {4\pi}} \right)
               G^\ast \delta \Phi ~~.
\ee$$
We have found that the slopes of the $\beta$~functions around the fixed
point are negative definite.
We conclude that the fixed point is ultraviolet stable against all
possible perturbations.
{}From these considerations it appears likely that if we start with the
Einstein theory, the renormalization trajectory is attracted to this
ultraviolet stable fixed point as $\mu \rightarrow \infty$.

At the initial stage of the evolution, we may parametrize L and $\Phi$
as
$$L \sim 1 + \lambda \psi \quad ,$$
$$\Phi \sim \kappa \psi \quad ,\ee$$
where $\lambda$,$\kappa \sim \rm{O}(\sqrt\epsilon)$.
The $\beta$~functions are
$$\beta_G \sim \epsilon G,$$
$$\beta_\lambda \sim - {A\over 2} G \lambda,$$
$$\beta_\kappa \sim {A\over 2} G \kappa.\ee$$
We can see that $\lambda$ decreases while $\kappa$ increases during
the evolution at the initial stage.
Since $\lambda$ is $\rm{O}(\sqrt\epsilon)$ to start with and gets
smaller during the evolution at the initial stage, we may neglect it
altogether and consider the evolution of $\Phi$ only.
If we retain the linear part in $\Phi$ (namely $\kappa$ ), the
evolution of $\kappa$ obeys the simple renormalization group equation:
$$\mu {d\over {d\mu}}\kappa =
\beta_\kappa = \left( {A\over 2} \kappa -
                 {{c\kappa^3} \over {16\pi}} \right) G .
\ee$$
With these approximations, we can solve the renormalization group equation
analytically and find that the theory approaches the ultraviolet stable
fixed point of the $\beta$~functions as $\mu \rightarrow \infty$.
Although such an approximation makes the problem tractable and
presents us a clear physical picture, it is also possible to integrate
the renormalization group equations numerically without any other
approximations [16].
In this way it has been numerically checked that the renormalization
trajectory which starts with the Einstein action in the infrared limit
belong to the domain of the attraction of the ultraviolet stable fixed
point we have found.

Our conclusion from these investigations is that the quantum gravity
in $2+\epsilon$ dimensions possesses the ultraviolet fixed point
when the matter content of the theory is such that $0<c<25$.
When the coupling G is weak, the long distance property of the
space time is described by the Einstein theory with the action
$$
{{\mu^{\epsilon}}\over G} {\tilde R}
{\left( 1 + {1 \over 2} {\sqrt {\epsilon \over 2}} \psi \right)}^2
-{{\mu^{\epsilon}} \over {2G}} \partial_\mu \psi \partial^\mu \psi
+ {1\over 2} \sum_i \partial_\mu \varphi^i \partial^\mu \varphi^i
{\left(1 + {1\over 2} {\sqrt{\epsilon \over 2}} \psi \right)}^2.
\ee$$
However the short distance property of the space time is governed by
the ultraviolet stable fixed point with the following action which
incorporates the renormalization effect due to quantum fluctuations of
the space time :
$$
{{\mu^{\epsilon}} \over {G^\ast}} {\tilde R} -
{\mu^\epsilon \over {2G^\ast}} \partial_\mu \psi \partial^\mu \psi
+ \sum_i {1\over 2} \partial_\mu \varphi^i \partial^\mu \varphi^i
e^{\Phi^{\ast(\psi)}} \quad .
\ee$$

\chapter{Renormalization of composite operators}

In this section we study the gravitational dressing problem of
spinless operators with scaling dimension $2\Delta_0$ denoted by
$\Phi_{2\Delta_0}$. From the general covariance it is
gravitationally dressed in the weak coupling regime as
$$
\int d^D x \sqrt{g} \Phi _{2\Delta_0}(g).\ee
$$
However the operator $\Phi_{2\Delta_0} $ also depends on the metric.
Its dependence is determined by the scaling dimension as
$\sqrt{g}^{-{2 \over D}\Delta_0}$.
We may consider $e^{i {\rm k} \varphi}$ as an example. $\varphi$ is
a free field with the kinetic term as
${1 \over 2}e^{-{\varepsilon \over 2}
\phi}{\tilde g}^{\mu\nu} \partial_\mu\varphi \partial_\nu \varphi  $
where the conformal mode dependence is explicitly shown. In order
to make the composite operator $e^{i {\rm k} \varphi}$ finite, we
need to subtract the divergence as
$$
\eqalign{
& Z < e^{i {\rm k} \varphi} > \cr
& \sim e ^{- {{\rm k}^2 \over{4\pi\epsilon}}}e^{{1 \over{4\pi\epsilon}}
{\rm k}^2 e^{ {\epsilon \over 2} \phi}} \quad , \cr
& \sim e^{{{{\rm k}^2} \over {8\pi}} \phi}.
}\ee$$
Since the dimension of this vertex operator is
$2\Delta_0 ={{\rm k}^2 \over {4\pi}}$,
the conformal mode dependence of the normal ordered operator
is  $e^{{{\rm k}^2 \over{8\pi}} \phi} =
{ \sqrt {g}}^{-{ 2 \over D} \Delta_0} $.
Therefore we need to investigate the renormalization of the composite
operator $\sqrt{g}^{1-{2 \over D }\Delta_0}$ when we study the
gravitational dressing of the operator $ \Phi_{2\Delta_0}$.

In the following, we study the renormalization of
the composite operator $f(\psi)$ in
$2+\epsilon$ dimensional quantum gravity
around the ultraviolet fixed point.
The strategy to study the theory around the ultraviolet fixed point
has been explained in sections 3 and 4.

The bare composite operator corresponding to $f(\psi)$ at the one-loop
level is 
$$
\eqalign{f_0(\psi) & =f(\psi)- {G\over {4\pi\epsilon}}f''(\psi) \quad ,
 \cr
&= f\left( \psi_0 + G \int_0^{\psi_0} dy
    \left\{{c\over{16\pi\epsilon}}{\left(\Phi'(y)\right)}^2
    -{A\over {2\epsilon}} - {{L''(0)}\over{8\pi\epsilon}}
                                                  \right\} \right) \cr
& -{G\over {4\pi\epsilon}}f''(\psi_0) \quad . \cr
}\ee$$
Therefore the $\gamma$~function for $f(\psi)$ is given by
$$
\gamma_{{}_f} \equiv \mu{d\over{d\mu}}f(\psi) =
               {G\over {4\pi}} f''-f'\gamma_\psi \quad .\ee
$$
We introduce $\tilde\psi$ which is a solution of
$$
\mu{d\over{d\mu}} {\tilde\psi} = -\gamma_{\tilde\psi} \quad , \ee
$$
where $\gamma_{{}_{\tilde\psi}}$ is obtained from $\gamma_\psi$ in
eq.$(4.5)$ by replacing $\psi$ by $\tilde\psi$ .
We also consider a new function $F(\tilde\psi) = f(\psi)$.
Then the $\gamma$~function for $F(\psi)$ simplifies as
$$
\mu{d\over {d\mu}} F(\tilde\psi) =
                         {G\over{4\pi}} F''(\tilde\psi) \quad. \ee
$$
If we approximate $\Phi\sim\kappa\psi$ and $L\sim 1$ we obtain
$$
\tilde\psi = \kappa\psi \ee
$$
where $\kappa$ is the solution of $(4.12)$ .

We seek a perturbative solution of (5.6) by using the approximation
$\mu{\partial\over{\partial\mu}}G=\epsilon G$ which is valid in the
weak coupling regime.
The solution of this diffusion equation is
$$
F(\kappa {\psi}) = \int^{\infty}_{-\infty} d\psi'
{\sqrt{\epsilon\over G}} {\rm exp} \left(-{{\pi\epsilon}\over G}
({\psi} - \psi')^2 \right) F_I(\kappa \psi').\ee
$$
Let us recall that the weak coupling limit of $\kappa$ is
$\kappa = \sqrt {\epsilon \over 2}$.
Hence we may take
$$
\eqalign{F_I(\kappa \psi)
        & = {\rm exp}\left\{ {4\over\epsilon} (1-\Delta_0){\rm log}
                                 (1+{1\over2}\sqrt{\epsilon\over2}
            \psi)\right\}~~,    \cr
        & = {\rm exp}\left\{ {4\over\epsilon} (1- \Delta_0){\rm log}
                        (1+{1\over2}\kappa \psi)\right\}.
            \cr
            }\ee
$$
The integral (5.8) can be evaluated by using the saddle point
approximation when $\epsilon \sim 0$ and $G \kappa^{2} \gg \epsilon^2$
$$
\eqalign{F(\kappa {\psi})
         &  =\int^{\infty}_{-\infty}d\rho {\rm exp}
                                 \left\{ -{{4\pi\epsilon}
             \over{G\kappa^2}}({\kappa{\psi}\over2}-\rho)^2 +
            {4\over \epsilon}(1-{\Delta_0})\log(1+\rho)\right\}~~,\cr
         &  \sim {\rm exp} \left( {4\pi\epsilon \over {G{\kappa^2}}}
            \rho_{{}_0} \kappa \psi \right)~~,                  \cr
             }\ee$$
where $\rho_{{}_0}$ is
$$
\rho_{{}_0}={1\over2}
 \left\{ -1 +
    \sqrt{1+ {2G{\kappa^2}\over{\pi\epsilon^{2}}}}
                                          \right\}.\ee
$$

If we naively use these results at the fixed point where
${G^\ast}= {\epsilon \over A}$ and  $ {(\kappa^2)^{\ast}}
= {8\pi A \over c}$, we have
$$
\eqalign{ \rho^{\ast}
           & ={1\over2} \left\{-1 + \sqrt{1 + {16({1-\Delta_0}) \over c
              \epsilon}} \right\}\quad ,   \cr
           & =2{\sqrt {1-{\Delta_0} \over c\epsilon}}
              \left(1- {1\over4}\sqrt{c\epsilon \over {1-\Delta_0}}
              + \cdot \cdot \right),   \cr
}\ee$$
$$
F^{\ast} (\kappa \psi) = {\rm exp} \left\{ {1\over {G^\ast}}
                 {\sqrt{{8\pi(1-{\Delta_0})\epsilon} \over A}}
        \left(1- {1\over4}{\sqrt{c\epsilon \over {1-\Delta_0}}}\right)
        \psi \right\} .\ee
$$
The scaling dimension of the gravitationally dressed operator
$  \int d^D x \sqrt{\hat g}^{1-\Delta_0} F^{\ast}(\kappa \psi) $
would be
$$
- \root \of{c \epsilon(1- {\Delta_0})}\ee
$$
at the ultraviolet fixed point.

However we need to examine the validity of our calculation near the
ultraviolet fixed point more closely.
If we use the one-loop $\beta_G$ function in (4.5) instead of the
tree $\beta_G=\epsilon G$ in our calculation, we find that the
renormalized operators behave differently near the ultraviolet fixed
point.
Although they become nearly marginal operators at the intermediate
stage, the anomalous dimensions are found to vanish eventually at the
ultraviolet fixed point.
Since we are dealing with the operators which possess $1\over \epsilon$
singularities to start with , we may use the following equation
$$
\mu {\partial\over {\partial\mu}} F(\kappa\psi) =
    {{\beta_G} \over {4\pi\epsilon}} F''(\kappa\psi) \ee
$$
instead of eq.(5.6).
If we do so, we retain the results of our perturbative calculation.
Furthermore the higher order corrections of the form
$G^{(n-1)} F^{(n)} (\kappa\psi)$ may not be negligible since $\kappa$
grows up
to $O(1)$.

Due to the presence of these uncertainties, we resort to the conformal
invariance arguments to determine the scaling dimensions of the
gravitationally dressed operators at the ultraviolet fixed point.
As it will be shown below, our perturbative results are
justified after such considerations.

For this purpose we recall the following nonlinear sigma model
Lagrangian [17-19]:
$$
{\cal L} = {1\over {2G}} {\sqrt g} g^{\mu\nu} G_{ij}(X)
\partial_\mu X^i \partial_\nu X^j
+ {1\over G} {\sqrt g} L(X) R^{(2)} \quad ,
\ee
$$
where $G_{ij}$ is the $N$ dimensional target space metric.
$G$ is the loop expansion parameter which is introduced in order to
compare eq.(5.16) with eq.(3.1).
In the conformal gauge, $g_{\mu\nu}=e^{-\rho}\delta _{\mu\nu}$
and we need to add the standard ghost Lagrangian.
The renormalization group $\beta$ functions are
$$\baselineskip 1.2cm
\eqalign{
\beta^G_{ij} &= - {1\over{2\pi}} R_{ij} + \cdots \quad ,\cr
\beta^L& = {{26-N}\over {24\pi}} - {1\over {4\pi}}
                                  D^i \partial_i L + \cdots \quad .\cr
}\ee$$
These quantities are in general different from the Weyl anomaly
coefficients ${\bar\beta}^\alpha$ which are defined by
$$\baselineskip 1.2cm
\eqalign{
{{\bar\beta}^G}_{ij} &= \beta^G_{ij} - {4\over G} D_i \partial_j L
\quad , \cr
{\bar\beta}^L &= \beta^L - {2\over G} \partial^i L \partial_i L  \quad , \cr
}\ee$$
where the difference comes from the tree level breaking of the
conformal invariance.
This can be seen by considering the transformation
$g_{\mu\nu} \rightarrow e^{-\rho} g_{\mu\nu} \quad
X^i \rightarrow X^i-(\partial^iL)\rho $ .

If we generalize this approach to $2+\epsilon$ dimensions,
we may expect
$$\baselineskip 1.2cm
\eqalign{
\beta^G _{ij}&= -{\epsilon \over G} G_{ij}-{1\over{2\pi}}
                                 R_{ij} + \cdots \quad , \cr
\beta^L &= -{\epsilon \over G} L + {{26-N}\over {24\pi}}
           -{1\over {4\pi}} \partial^i \partial_i L + \cdots \quad.\cr
}\ee$$
One of the fixed points of these $\beta$ functions corresponds to the
de Sitter space with the metric
$$
{\sqrt g} g^{\mu\nu} G_{ij}^* \partial_\mu X^i \partial_\nu X^j
 = - {\sqrt g} g^{\mu\nu} \partial_\mu \psi \partial_\nu \psi +
   \sum_{i=1}^c {\sqrt g} g^{\mu\nu} {\rm exp}
   \left({\sqrt{{25 - c}\over {3c}}} \psi \right)
\partial_\mu \varphi^i \partial_\nu \varphi^i \ee
$$
and with
$$
L^* = 1,\quad\quad  G^* = {{24 \pi \epsilon } \over {25 - c}} \quad ,
\ee$$
where
$$
c=N-1 \quad. \ee
$$
Note that this is nothing but our fixed point (4.7).
Furthermore the only contribution from the ghost sector is
$26\over{24\pi}$ to $\beta^L$ just like that from $h_{\mu\nu}$ and
ghost fields in our gauge.
Therefore our result is consistent with a naive extension of the
conformal gauge approach to string theory.
Since we have chosen a different gauge, this observation implies the
one-loop universality of the $\beta$ functions.

In the nonlinear sigma model approach to string theory,
the conformal invariance plays a crucial role.
Let us briefly summarize the reasoning.
We decompose the metric $g_{\mu\nu}$ into the background metric
${\hat g}_{\mu\nu}$ and the quantum fluctuations.
The theory must be invariant under the conformal transformation of the
background metric ${\hat g}_{\mu\nu} \rightarrow {\hat g}_{\mu\nu}
e^{-\rho}$ since we are integrating the quantum fluctuations which
include the conformal mode.
This requirement is very general for quantum gravity, and we expect it
to hold in any dimension.

In particular the quantum gravity in $2+\epsilon$ dimensions at the
ultraviolet fixed point is certainly conformally invariant since the
Weyl anomaly (5.18) vanishes there.
Therefore we apply the standard strategy of the two dimensional gravity
at the ultraviolet fixed point.
Namely, we require that the physical operators in the theory are
conformally invariant.

Since $\psi$ decouples from $\tilde R$ at the ultraviolet fixed point,
the gravitationally dressed operators are determined by the conformal
invariance to be
$$
\int d^D x \sqrt{\hat g} \Phi_{2\Delta_0} e^{\beta\psi} \quad ,\ee
$$
where $\Delta_0 + {{G^* \beta^2} \over {8\pi}}=1$ .
Here we have neglected the interactions among $\psi$ and $\varphi^i$
fields since they are $O(\sqrt\epsilon)$.
In particular the cosmological constant operator is
$$
\int d^D x \sqrt{\hat g} e^{\alpha\psi} \quad . \ee
$$
The theory at the ultraviolet fixed point is invariant if we shift
$$
\eqalign{
\psi \rightarrow & \psi+{\lambda\over\alpha} \quad ,\cr
\varphi^i \rightarrow &
e^{-{1\over 2}\sqrt{{25-c}\over 3c} {\lambda\over\alpha}} \varphi^i =
{(\mu')}^{{1\over 2}\sqrt{\epsilon\over c}} \varphi^i  \quad , \cr
}
\ee$$
where the cosmological constant operator is used as the standard of the scale.
Namely, the cosmological constant operator scales as
$$
\int d^D x \sqrt{\hat g} e^{\alpha\psi} \rightarrow
          (\mu')^{-1} \int d^D x \sqrt{\hat g} e^{\alpha\psi} \quad ,
\ee$$
and the general gravitationally dressed operators scale as
$$
\int d^D x \sqrt{\hat g} \Phi_{2\Delta_0} e^{\beta\psi} \rightarrow
(\mu')^{-\sqrt{1-\Delta_0}}\int d^D x \sqrt{\hat g}
\Phi_{2\Delta_0} e^{\beta\psi}\quad .
\ee$$
On the other hand from $\gamma_{\varphi^i}$ in(4.5),
we also know that the anomalous dimension of $\varphi^i$
vanishes at the ultraviolet fixed point.
Therefore we conclude that $\mu' = \mu^{\sqrt{c\epsilon}}$ since
$\varphi^i$ must scale canonically
$\varphi^i \rightarrow \mu^{\epsilon\over 2}\varphi^i $
at the ultraviolet fixed point.

After these considerations we find that the scaling dimensions of the
gravitationally dressed operators
$\int d^D x \Phi_{2\Delta_0}e^{\beta\psi}$
are indeed
$$-\sqrt{c\epsilon(1-\Delta_0)} \ee$$
to the leading order of
$\epsilon$, which agree with our perturbative results.
In particular the scaling dimension of the cosmological constant
operator is
$
-\sqrt{c \epsilon}.
$
Hence in the limit
$\epsilon \to 0 $
, the cosmological constant operator and all other gravitationally
dressed operators become marginal operators at the ultraviolet fixed
point.
Strictly speaking the conformal invariance cannot determine the signs
of $\alpha$ and $\beta$.
Therefore we cannot exclude the possibility that the irrelevant
operators which correspond to the other branch of $\alpha$ and
$\beta$ may exit in the theory.
In the weak coupling limit, the cosmological constant operator
$\sqrt g $ is a dimension zero operator. Due to the quantum
fluctuations, at the short distance the cosmological constant operator
acquires the anomalous dimension and becomes nearly marginal.
However it is still the most relevant operator in the theory as long as
$ \epsilon\neq 0$.
There is another relevant coupling $ G $ in the theory since
$ \beta_G = - \epsilon \delta G$ near the ultraviolet fixed point
(see eq.(4.9)).

We may observe that the gravitational coupling $ G $ is analogous to
$ T - T_c$ in the spin system (nonlinear sigma models or scalar
field theory).
On the other hand the cosmological constant is analogous to the
magnetic field of the spin systems.
Just like nonlinear sigma models, we have two
relevant operators which should be tuned around the ultraviolet fixed
point in $ 2 + \epsilon $ dimensional quantum gravity.

Namely with the renormalization scale $\mu$, the gravitational
coupling and the cosmological constant scale as
$$
\delta G \sim \mu^{-\epsilon} \quad ,
$$
$$
\Lambda \sim \mu^{-\sqrt{c\epsilon}},\ee
$$
so that
$$
\Lambda \sim {(G-G^\ast)}^{\sqrt{c\over \epsilon}}~. \ee
$$
We may plot the renormalization group flow of the couplings on the 2
dimensional plane in Fig.1.

\goodbreak\midinsert
\figin{\centerline{\epsfysize=2.5in \epsfbox{fig1.eps}}}
\centerline{\tenpoint Fig.1 ~ The transverse direction is $G$ and the
vertical direction is $\Lambda$.}
\endinsert
As we decrease the renormalization scale  $\mu$ , the flow of the
coupling is shown by lines with an arrow.  Since $\Lambda$ controls
the size of the universe, we predict that two universes with
different volume (cosmological constant) and different gravitational
coupling constant look identical if they are related by the
relation (5.30).

Therefore (5.30) is a scaling relation of the quantum gravity in
$ 2 + \epsilon $ dimensions.  The derivation of such scaling
relations is very important for understanding quantum gravity.
In the constructive approach to quantum gravity such as the dynamical
triangulation method, we need to understand such scaling relations
to search for the continuum limit.  In fact it has not been made
clear what one should observe in such an approach in order to
search for the continuum limit.

Better understanding of quantum gravity should derive scaling
relations which are relevant to continuum limit of the constructive
approach and eq.(5.30) is such an example.

\chapter{Two dimensional gravity and string theory in our framework}

Although we are mostly interested in the quantum gravity beyond
two dimensions in this paper, our approach provides new insights and possibly
new
results in tow dimensional gravity and string theory. In this section we would
like to discuss the renormalization property of the quantum gravity with the
positive coupling $G$ in $2-\epsilon$
dimensions briefly in connection with the two dimensional quantum
gravity. As before we consider the most general renormalizable action
$$
{{\mu^{-\epsilon}} \over G} {\tilde R} L(\psi) +
{1\over 2} {{\mu^{-\epsilon}}\over G}
\partial_\mu \psi \partial^\mu \psi +
{1\over 2} \partial_\mu \varphi^i \partial^\mu \varphi^i
e^{\Phi(\psi)}.
\ee$$
Note that the sign of the kinetic term of the conformal mode is
positive in $2-\epsilon$ dimensions.
The one-loop $\beta$~functions are
$$
\beta_G = - \epsilon G - A G^2 + {{L''(0)} \over {4\pi}} G^2,
$$
$$
\beta_L  = GL'(\psi) \int_0^\psi dy
 {\left\{ {c \over 16} {\left( \Phi'(y) \right)}^2
         + {A \over 2} - {{L''(0)} \over {8\pi}} \right\}}
$$
$$
+ \left\{A - {{L''(\psi)} \over {4\pi}} -AL(\psi) +
     {{L''(0)} \over {4\pi}}  L (\psi)\right\}G \quad ,
$$
$$
\beta_\Phi = G \Phi'(\psi) \int_0^\psi dy
 {\left\{ {c \over {16\pi}} {\left(\Phi ' (y) \right)}^2
 + {A \over 2} - {{L''(0)} \over {8\pi}} \right\}}
$$
$$
+ \left\{- {1 \over{4\pi}} \Phi''(\psi) + {1 \over {4 \pi}} \Phi''(0)
\right\} G \quad.
\ee$$
G gets weaker as the renormalization scale $\mu$ increases.
Therefore we may start with the Einstein theory in the ultraviolet
limit $(\mu \rightarrow \infty)$ and ask what happens when the
renormalization scale $\mu$ is decreased.
In this case, G and L grow while $\Phi$ diminishes during the
evolution in contrary to the previous situation.

If we consider the scale where G is large compared to $\epsilon$
but still small such that we can trust the perturbation theory
($\rm{O} (\sqrt \epsilon)$ to be precise), the effect of L to the
renormalization group turns out to be still small.
It is numerically found [16] that
$L''(0) \sim \rm{O}
(\epsilon^{3\over 5})$.
Let $\Phi \sim \kappa \psi$ as in section 5.
$\beta_\kappa$ is well approximated by
$\beta_\kappa \sim {A\over 2} \kappa G$.
We remark that $\kappa$ determines the scale of $\psi$ field.
As it was shown in section 5 that any operator
$f(\psi)$ is renormalized
in a definite combination of $\kappa$ and $\psi$ as $F(\kappa \psi)$.
In other words, the effective propagator of $\psi$ is proportional
to $G\kappa^2$.
Note that
$$
\mu {\partial \over {\partial\mu}}
\left( {1 \over {G \kappa^2}} \right)
= \epsilon \left({1 \over {G \kappa^2}} \right).
\ee$$
Therefore the effective inverse propagator ${1\over k'}$ of
$\psi$ scales
canonically and it is found to be
$${1\over k'}={1 \over 2} \left( {{25-c} \over {24\pi}}
                  + {\epsilon \over G} \right) \quad ,
\ee$$
by solving the renormalization group with the initial condition
$\epsilon \over {2G}$ when $\mu \rightarrow \infty$.
Here we have scaled back
$\psi \rightarrow \sqrt{\epsilon \over 2} \psi$.
When $G \sim \rm{O}(\sqrt\epsilon)$, the inverse propagator is equal
to ${{25-c} \over {48\pi}} + \rm{O}(\sqrt\epsilon)$.
In the previous work, we have shown that such a theory reproduces
the exact scaling exponents of two dimensional quantum gravity in the
limit $\epsilon \rightarrow 0$.

In order to investigate the gravitational dressing of spinless
operators $\Phi_{2\Delta_0}$ with scaling dimension $2\Delta_0$,
we consider the renormalization
of a most general composite operator $ F(\psi) $.
$\psi$ is the redefined conformal mode whose kinetic term is canonical:
$$
{1\over{2 \kappa'}}\mu^{-\epsilon}\partial_\mu \psi \partial^\mu \psi.
\ee$$
where $1\over {k'}$ is given in (6.4).
It is found that the vacuum expectation value of the following
bare operator is finite:
$$
F_0 (\psi) = e^{-{\kappa' \over {4\pi \epsilon}}{\partial^2 \over
{\partial \psi^2}}}F(\psi). \ee$$
The renormalization group for $F(\psi)$ follows as
$$
\mu{\partial \over \partial\mu} F (\psi)= - {\kappa'\over{4\pi}}
{{\partial^2 F(\psi)} \over {\partial \psi^2}}
 \equiv - {\kappa' \over {4\pi}}  F'' (\psi) \quad .\ee
$$
In these derivations we have used the relation
$ \mu {\partial \over \partial\mu}\kappa' = - \epsilon \kappa'$.
Equivalently we may change the variable from $\mu$ to $\kappa'$ as
$$
4\pi \epsilon {{\partial F (\psi)} \over {\partial \kappa'}} =
                                                    F''(\psi).\ee
$$
Therefore we conclude that the renormalized operator $F(\psi)$
satisfies the diffusion equation.
The solution of this diffusion equation is
$$
F(\psi)= \int^\infty_{-\infty} d\psi' \sqrt{2 \over \kappa'}
{\rm exp}(- {{\pi \epsilon} \over \kappa'} ({\psi} - \psi')^2)
F_I(\psi') \quad ,\ee
$$
where the initial condition $F_I(\psi)$ needs to be specified.

In our diffusion equation, the coupling $\kappa'$ play the
role of time. We consider such a situation that at the beginning
$\kappa'$ is small and it grows later.  This kind of situation
occurs in $ 2-\epsilon $
dimensions if we decrease the renormalization scale $ \mu $ .
In other words the theory goes into the strong coupling region from
the weak coupling region.
In the weak coupling region, the composite operators are not
renormalized and $F_I(\psi)$ can be taken to be
${\sqrt g}^{1-\Delta_0}$ :
$$
F_I(\psi) =
{\rm exp}\left\{ -{4 \over \epsilon} (1 - \Delta_0)\log
            (1 - {\epsilon \over 4}\psi)\right\}.
\ee
$$
With this initial condition, we can solve the diffusion equation and
find the form of the renormalized operators when the coupling
becomes strong.
When $ \epsilon \sim 0$, we can use the saddle point
method to find the following renormalized operator :
$$
\int^{\infty}_{-\infty}d \rho {\sqrt{{Q^2}\over{8\pi}}} {\rm exp}
\left\{ -{{Q^2}\over{\epsilon}} (\rho + {\epsilon\over4} \psi)^2
-{4\over\epsilon} (1-{\Delta_0})\log(1+\rho) \right\},
\ee
$$
where from eq.$(6.4)$
${1\over k'}\sim{Q^2\over16\pi}$ when ${G \gg \epsilon}$.
Eq.(6.11) is evaluated to be
$$
{\left(Z_{\Delta_0}\right)}^{-1} {\rm exp}
             \left(-{Q\over2} \beta {\bar \psi}\right),\ee
$$
where
$$\beta=-{Q\over2}\left\{ 1 - \sqrt{1- {{8(1-\Delta_0)}\over Q^2}}
\right\}.\ee
$$

In sect.2, we have explained that the conformal mode is governed by the
following Liouville action when $G\gg\epsilon$ :
$$
{1\over 8\pi} {\big({Q\over2}\big)^2} \int d^D x \sqrt{\hat g}
(\tilde g^{\mu\nu} \partial_{\mu}\psi\partial_{\nu}\psi
- 2\psi{\tilde R}).\ee
$$
In this section we have further derived the renormalized operators
of the form

$$\int d^D x e^{{-{Q\over2}}\beta\psi} \Phi_{2\Delta_0}.\ee$$

By rescaling the conformal mode as $\psi \to {Q\over2}\psi$,
we have found that
these results are in complete agreement with the conformal gauge
approach in two dimensional gravity. Therefore we have given the
renormalization group justification of the conformal gauge approach to
two dimensional gravity. The scaling exponents of the gravitationally
dressed operators are the sum of the canonical and the anomalous
dimensions:

$$
2(1-\Delta_0) + \mu{\partial\over{\partial\mu}}{\rm log}
                                     {Z_{\Delta_0}}=-Q\beta .\ee
$$

The string susceptibility of the theory in the two dimensional limit
can be derived by scaling the metric
$g_{\mu\nu} \to \lambda g_{\mu\nu}$
in the microcanonical ensemble:

$$\eqalign{ Z
    & =\int Dg_{\mu\nu} e^{-\int\lambda^{-{\epsilon\over2}}{1\over{G_0}}
       {\sqrt g}R}
       \delta \left(\lambda^{D\over2}<\int{\sqrt g}
       >\Bigl|_{\mu\lambda^{1\over2}}- A \right) \cr
    & =e^{{\epsilon\over2}{Q^2\over{8\pi\epsilon}}\log \lambda
                              \int {\sqrt g}R}
       \int Dg_{\mu\nu}e^{-{\int{1\over {G_0}}{\sqrt g}R}}
       \delta \left(\lambda^{{\alpha\over2}(-Q)} <\int {\sqrt g}>
       \Bigl|_{\mu} - A \right)  \cr
    & =A^{-1+{Q\over\alpha}} \int Dg_{\mu\nu}e^{-\int{1\over G_0}
       {\sqrt g}R} \delta (<\int {\sqrt g}>\Bigl|_{\mu}- 1),  \cr
        }\ee$$
where $\int{\sqrt g}R = -8\pi $ on two dimensional sphere is used.
The string susceptibility $\Gamma = 2 + {Q\over \alpha}$ is reproduced
in the two dimensional limit. Needless to say, these renormalization group
considerations are in accord with the arguments which use the shift of
the constant mode of $\psi$.

As for the ultraviolet fixed point described in $2+ \epsilon$
dimensions it may have applications to string theory.
We point out that our ultraviolet fixed point is very similar to the critical
string in
the sense that conformal mode becomes a free field when
$\epsilon \sim 0$ .
One might hope that $\epsilon \sim 0$ limit of the fixed point leads
to a new string theory with $c~>~1$.
However such a theory turns out to not differ much from the
conventional
string theory except that the cosmological constant operator is made
of a free field (conformal mode).

As it can be seen from the first term of $\beta_L$ in eq.(5.19)
the gravity sector has contributed to the central
charge as ${\epsilon \over G^*} \propto 25-c$ in the
$\epsilon \rightarrow 0$ limit.
Such a theory may be represented by
Brans-Dicke type gravity in two dimensions.

\chapter{Conclusions and Discussion}

In this paper we have proposed a formulation of renormalizable quantum
gravity in $2+\epsilon$ dimensions.
Our formulation can be regarded as a natural extension of the
nonlinear sigma model approach to string theory (two dimensional
gravity)[17-19] into $2+\epsilon$ dimensions.

We have generalized the Einstein action by relaxing the conformal mode
dependence
$$
\int d^D x \left({1\over G} {\sqrt {\tilde g}} {\tilde R} L(\psi) -
 {1\over {2G}}{\sqrt {\tilde g}}{\tilde g}^{\mu\nu}
    \partial_\mu \psi \partial_\nu \psi +
{1\over 2} e^{\Phi(\psi)} \sqrt{\tilde g} {\tilde g}^{\mu\nu}
    \partial_\mu \varphi^i \partial_\nu \varphi^i \right). \ee
$$
Under the general coordinate transformation,
$
{\tilde g}_{\mu\nu} = {\left( {\hat g} e^h \right)}_{\mu\nu}
$
transforms as
$$
{{\tilde g}_{\mu\nu}}'(x)={{\partial{x'}^\rho}\over {\partial x^\mu}}
                        {{\partial{x'}^\sigma} \over {\partial x^\nu}}
                        {\tilde g}_{\rho\sigma} (x')
\left(1-{2\over D} {\hat D}_\tau \epsilon^\tau \right).
\ee$$
Namely it does not transform as a tensor but it undergoes extra
conformal transformation.
Since the action (7.1) is invariant if ${\tilde g}_{\mu\nu}$ transforms as a
tensor,
it breaks the general covariance if it is not
conformally invariant.

However it is invariant under a restricted class of the
general coordinate transformation such that
${\hat D}_\rho \epsilon^\rho=0$
(volume preserving diffeomorphism).
Furthermore it is invariant under the constant mode
shift of $\varphi^i$.

Hence the one-loop matter contributions are renormalizable within this
action.
However we need to add the gauge fixing and the ghost terms to quantize
the gravity sector.
In doing so we have no other choice than to add BRS trivial terms with
respect to the general coordinate transformation.
Since the action (7.1) is not BRS invariant, we may worry whether the
theory is really renormalizable within the action (7.1) plus BRS
trivial terms.
However we have already seen a lot of cancellations in gravity sector.
As was shown in section 3, the action (7.1) is renormalizable to the
one-loop order.
Since we can choose the conformal gauge in two dimensions where
$h_{\mu\nu} = 0$ , we may expect that the divergences due to
$h_{\mu\nu}$ field are down by single power of $\epsilon$.
Even if it is the case, we cannot exclude the possibility that
divergences arise which cannot be renormalized within the action (7.1)
in higher orders.

If so, we may need to introduce all possible renormalizable terms in
addition to the action (7.1).
The theory is certainly renormalizable if we do so, but it may become
computationally more difficult to deal with.
Nevertheless these additional couplings are expected to be small since
the theory is nearly conformally invariant in $2+\epsilon$ dimensions.

Another feature of the action (7.1) is the negative sign of the
kinetic term of the conformal mode.
In two dimensions the sign of the kinetic term of the conformal mode
is positive as long as $c < 25$, because the theory is in the strong
coupling phase and it is far from the Einstein gravity.
On the other hand beyond two dimensions, the sign of the kinetic term of
the conformal mode is inevitably negative in the weak coupling regime.
Unlike in two dmensions, the simple map to the statistical systems
beyond two dimensions may not work in the weak coupling region due to
the instability of the conformal mode.
This is the inevitable problem in realistic quantum gravity and
further considerations may be needed in the dynamical triangulation
approach with respect to this problem.

We have again seen the well-known connection between the
conformal invariance and the BRS invariance in quantum gravity near
two dimensions in these considerations.
In the nonlinear sigma model approach to string theory, all possible
renormalizable couplings which break BRS
invariance in general are considered.
The BRS invariance can be imposed later by demanding the conformal
invariance.
We are proposing a similar strategy in $2+\epsilon$ dimensional
quantum gravity.
We consider all possible renormalizable  couplings which break BRS
invariance in general for the sake of the renormalizability.
We propose to impose the general covariance by picking up the
renormalization group trajectory which leads to the Einstein theory
in the weak coupling limit.

In this paper we have performed this program to the one-loop order.
We also have studied the renormalization of the generally covariant
operators.
Remarkably, the theory turns out to possess the ultraviolet fixed point
as long as $0<c<25$.

Furthermore we have computed the scaling dimensions of the relevant
operators in the theory near the ultraviolet fixed point.
Our results appear to be very encouraging toward the construction of
consistent quantum gravity in $2+\epsilon$ dimensions.
We hope to systematically improve our predictions in the near future.

\bigskip
\centerline{\fourteenbf Acknowledgements}
\smallskip

One of us (Y.K.) thanks T. Aida, T. Inui and S. Kojima for discussions
on this subject. we are grateful to Y. Okamoto for his careful reading
the manuscript. This work is supported in part by the Grant-in Aid for
Scientific Research from the Ministry of Education, Science and
Culture.

\endpage
{\fourteenbf References}

\item{[1]}E. Br$\acute{\rm e}$zin and V.A. Kazakov, Phys. Lett.
\undertext{236B}~ (1990)144.
\item{[2]}M. Douglas and S. Shenker, Nucl. Phys. \undertext{B335}~ (1990)635.
\item{[3]}D.J. Gross and A.A. Migdal, Phys. Rev. Lett. \undertext{64} (1990)
127.
\item{[4]}S. Weinberg, in General Relativity, an Einstein Centenary
Survay, eds. S.W.\nextline
Hawking and W. Israel (Cambridge University Press, 1979)p.790.
\item{[5]}R. Gastmans. R. Kallosh and C. Truffin, Nucl. Phys.
\undertext{B133}~(1978) 417.
\item{[6]}S.M. Christensen and M.J. Duff, Phys. Lett. \undertext{79B}~
(1978) 213.
\item{[7]}H. Kawai and M. Ninomiya, Nucl. Phys. \undertext{B336}~
(1990) 115.
\item{[8]}V.G. Knizhnik, A.M. Polyakov and A.A. Zamolodchickov,
Mod. Phys. Lett. \undertext{A3}~ (1988) 819.
\item{[9]}F. David, Mod. Phys. Lett. \undertext{A3}~ (1988) 651.
\item{[10]}J. Distler and H. Kawai, Nucl. Phys. \undertext{B321}~
(1989) 509.
\item{[11]}M.E. Agishtein and A.A. Migdal, PUPT-1287~ (1991).
\item{[12]}J. Ambj\o rn and S. Versted, NBI-HE-91-45~ (1991).
\item{[13]}M.E. Agishtein and A.A. Migdal, PUPT-1311~ (1992).
\item{[14]}H. Kawai, Y. Kitazawa and M. Ninomiya, UT-614, TIT/HEP
-191, YITP/U-92-05 (June 1992), to be published in Nucl. Phys.
\undertext{B}.
\item{[15]}G. 't Hooft and M. Veltman, Ann. Isnt. Henri Poincare
\undertext{20}~(1974) 69.
\item{[16]}T. Aida, Master Thesis at Tokyo Inst. Tech. (1993).
\item{[17]}C.G. Callan, D. Friedan, E.J. Martinec and M.J. Perry,
Nucl. Phys. \undertext{B262}~ \nextline
(1985) 593.
\item{[18]}E.S.Fradkin and A.A. Tseytlin, Nucl. Phys.
\undertext{B261}~(1985)~1.
\item{[19]}A.A. Tseytlin, Nucl. Phys. \undertext{B294}~(1987) 383.

\bye